# Ameliorate Threshold Distributed Energy Efficient Clustering Algorithm for Heterogeneous Wireless Sensor Networks

MOSTAFA BAGHOURI
Department of Physics
Faculty of Sciences
University of Abdelmalek Essaâdi
Tetouan, Morocco

SAAD CHAKKOR
Department of Physics
Faculty of Sciences
University of Abdelmalek Essaâdi
Tetouan, Morocco

ABDERRAHMANE HAJRAOUI
Department of Physics
Faculty of Sciences
University of Abdelmalek Essaâdi
Tetouan, Morocco

*Abstract*—Ameliorating the lifetime in heterogeneous wireless sensor network is an important task because the sensor nodes are limited in the resource energy. The best way to improve a WSN lifetime is the clustering based algorithms in which each cluster is managed by a leader called Cluster Head. Each other node must communicate with this CH to send the data sensing. The nearest base station nodes must also send their data to their leaders, this causes a loss of energy. In this paper, we propose a new approach to ameliorate a threshold distributed energy efficient clustering protocol for heterogeneous wireless sensor networks by excluding closest nodes to the base station in the clustering process. We show by simulation in MATLAB that the proposed approach increases obviously the number of the received packet messages and prolongs the lifetime of the network compared to TDEEC protocol.

## I. INTRODUCTION

Wireless sensor network is the set of sensor nodes, deployed in the hostile environment, in the goal to sense the events detection, such temperature, pressure or vibration and send their measurements toward a processing center called sink [1], [2]. These tiny nodes are limited in their battery capacity which its replacement is impossible. Furthermore, an important part of energy is consumed in the communication circuit which must be minimized. Because of those limitations, the major wireless sensor networks' challenging issues is the energy consumption.

A number of research techniques about energy-efficient have been proposed to solve these problems. In order to support data aggregation through efficient network organization, nodes can be partitioned into a number of small groups called clusters. Each cluster has a cluster head, and a number of member nodes [3]. Among WSN heterogeneous protocols there is DEEC (Design of a distributed energy-efficient clustering algorithm) [4]. This protocol is based on the election of cluster head by the balance of the remaining energy probabilities for each node. It uses the average energy of the network as the energy reference. The cluster-heads are elected by a probability based on the ratio between the residual energy of each node and the average energy of the network. DEEC has improved by a Stochastic approach SDEEC [5], which reduces the intra-cluster transmission. In this protocol the non-CH are going in to sleep mode to conserve more energy. Another version of improved DEEC is DDEEC which define a new residual energy threshold to elect CH [6]. On the other hand TDEEC enhance the network lifetime by introducing a new threshold based on the residual energy to become CH [7]. The last version of TDEEC is ETDEEC which prolong the lifetime by modifying the probabilities of CH election based on the distance average between the CHs and BS [8].

Otherwise, in order to improve the lifetime of the network, ATDEEC employs a new technique which excludes closest nodes to the base station from the clustering process. The remainder of the paper is organized as follows. In section II the main related works are summarized. Section III and IV introduced the problem formulation and proposed approach. Sections V and VI explains the network and the energy models. Therefore, theoretical analysis are presented and discussed in Section VII, whereas section VIII describes performance analysis of the proposed method. Finally, Section IX concludes our work, and discusses some future directions.

## II. RELATED WORK

Currently, clustered routing protocols have gained actually increasing attention from researchers because it's potential in extending WSN lifetime. Heinzelman et al. designed and implemented the first distributed and clustered routing protocol with low energy consumption LEACH [9]. Moreover, the heterogeneous protocols are more energy efficient than the homogeneous ones. Q. Li et al. have proposed Distributed Energy Efficient Clustering Protocol (DEEC) [4]. This protocol is based on multi-level and two level energy heterogeneous schemes. The cluster heads are selected using the probability utilizing the ratio between residual energy of each node and the average energy of the network. The epochs of being cluster-heads for nodes are different according to their initial and residual energy. A particular algorithm is used to estimate the network lifetime. Afterward, the network can avoid the need of assistance by routing protocol [4]. TDEEC [7] uses the same process of CH selection and estimation of average energy as in DEEC. At start of each round, the nodes decide whether or not to become a CH by selecting a random number within 0 and 1. If this selected number is lower than a threshold, then the node becomes a CH for this round. Simulation results show that in terms of network lifetime, both EDEEC and TDEEC protocols are better than DEEC. TDEEC provide best results compared to the three versions over DEEC. Otherwise, Suniti Dutt et al [6], has proposed ETDEEC protocol to enhance the network lifetime by introducing a distance factor in CH probability. However, this approach present a limitation lies in the fact that the network instability observed after the death of the first node is caused mainly by the bad energy distribution. It means that all nodes not die approximately at the same time.

### III. PROBLEM FORMULATION

In this paragraph we formulate the problem that we'll solve in the next sections. We consider a network with N nodes, which are uniformly distributed in a M × M network field as shown in Figure1.

Each node has a mission to send every time the data to the base station which is located at the center of network. This network divide in the cluster regions, and the cluster-heads receive the data from the member nodes to transmitting toward the base station. According to this model, it was found that the member nodes that are closer to the base station must go through a long path to route a data.

Contrariwise, they have the possibility to send the packet messages directly to the base station (Figure 1). In this case, these nodes should not go through the CH election process. Consequently we can conserve the lost energy during this step and we can prolong the network lifetime. To simulate this problem, we present in the next section the model of the studied network.

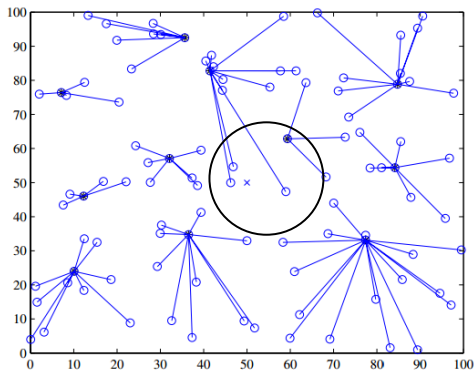

Fig. 1   Through the clustering process, all nodes must form clusters even those who are closest to the base station

### IV. PROPOSED METHOD

This paper proposes a new approach called Ameliorate Threshold Distributed Energy Efficient Clustering (ATDEEC) algorithm whose main objective is to increase the lifetime of the network and to enhance the ability to deliver more packet messages in the heterogeneous WSN by minimizing the number of the nodes elected to become CH.

### V. ENERGY MODEL

This study assumes a simple model for the radio hardware where the transmitter dissipates energy for running the radio electronics to transmit and amplify the signals, and the receiver runs the radio electronics for reception of signals [7]. Multipath fading model ($d^4$ power loss) for large distance transmissions and the free space model ($d^2$ power loss) for proximal transmissions are considered. Thus to transmit an $l-$ bits message over a distance d, the radio expends:

$$E_{Tx}(l,d) = E_{Tx-elec}(l) + E_{Tx-amp}(l,d) \quad (1)$$

$$E_{Tx-elec}(l) = lE_{elec} \quad (2)$$

$$E_{Tx-amp}(l,d) = \begin{cases} l\epsilon_{fs}d^2, when\ d < d_0 \\ l\epsilon_{mp}d^4, when\ d \geq d_0 \end{cases} \quad (3)$$

Where $d_o$ is the distance threshold for swapping amplification models, which can be calculated as $d_o = \sqrt{\frac{\epsilon_{fs}}{\epsilon_{mp}}}$

To receive an $l\ bits$ message the receiver expends:

$$E_{Rx}(l) = lE_{elec} \quad (4)$$

To aggregate n data signals of length $l-$ bits, the energy consumption was calculated as:

$$E_{DA-expend}(l) = lnE_{DA} \quad (5)$$

### VI. NETWORK MODEL

This section describes the network model and other basic assumptions:

*1) N sensors are uniformly distributed within a square field of area $A = M \times M$. The Base Station is positioned at the center of the square region. The number of sensor nodes N to be deployed depends specifically on the application.*

*2) All nodes are deployed randomly and can fall in the one of two types of regions which can be defined by the threshold distance R from the base station.*

*3) In this case we define two types of nodes, Excluded and not Excluded nodes. The Excluded are the nodes that not enter in the clustering process because there are closed to the base station and the other are far.*

*4) All sensors are heterogeneous, i.e., they not have the same capacities.*

*5) All the sensor nodes have a particular identifier (ID) allocated to them. Each cluster head coordinates the MAC and routing of packets within their clusters. (see Fig. 2)*

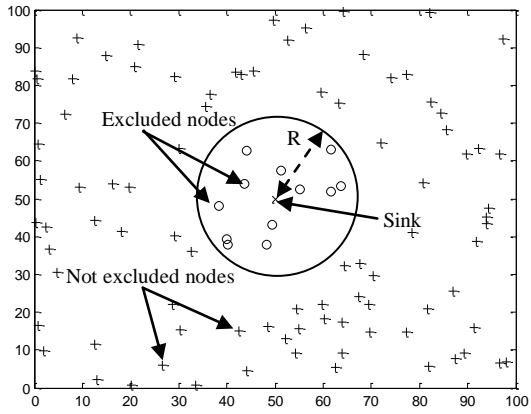

Fig. 2 Wireless Sensor Network model

## VII. THEORETICAL ANALYSIS

Let $E[d_{toBS\_ex}]$ be the Expected distance of Exclude node from the base station. Assuming that the nodes are uniformly distributed, so it is calculated as follows:

$$E[d_{toBS\_ex}^2] = \int_0^{x_{max}} \int_0^{y_{max}} (x^2 + y^2)\, \rho(x,y)\,dx\,dy \quad (6)$$

$$E[d_{toBS\_ex}^2] = \int_0^R \int_0^{2\pi} r^2 \rho(r,\theta)\,r\,dr\,d\theta \quad (7)$$

If the density of sensor nodes is uniform throughout the area then becomes independent of r and θ. It is equal to $\rho = \frac{1}{\pi R^2}$ then:

$$E[d_{toBS\_ex}^2] = \frac{R^2}{2} \quad (8)$$

According to the energy model proposed in section 5, the energy consumed by each Excluded nodes is:

$$E_{Exclu} = lE_{elec} + lE_{DA} + l\epsilon_{fs} d_{toBS\_ex}^2 \quad (9)$$

By combining the equations (8) and (9) the energy consumed by each Excluded nodes is:

$$E_{Exclu} = lE_{elec} + lE_{DA} + l\epsilon_{fs} \frac{R^2}{2} \quad (10)$$

The energy consumed by the Not Excluded nodes is:

$$E_{not\_Exclu} = cE_{cluster} = cE_{CH} + (N-s)E_{not\_CH} \quad (11)$$

Where $E_{CH}$ and $E_{not\_CH}$ are the energy consumed by each cluster head and member node respectively and can be calculated by:

$$E_{CH} = lE_{elec}\left(\frac{N-s}{c} - 1\right) + lE_{DA}\left(\frac{N-s}{c}\right) + lE_{elec} + l\epsilon_{mp} d_{toBS\_not\_ex}^4 \quad (12)$$

$$E_{not\_CH} = lE_{elec} + l\epsilon_{fs} d_{toCH}^2 \quad (13)$$

Where $d_{toBS\_not\_ex}$ is the average distance of not Excluded node from the base station and $d_{toCH}$ is the average distance between cluster members to CH.

Now $d_{toBS\_not\_ex}$ and $d_{toCH}$ can be calculated as:

$$d_{toBS\_not\_ex}^2 = \int_0^{\sqrt{\frac{M^2-\pi R^2}{\pi}}} \int_0^{2\pi} r^2 \rho(r,\theta)\,r\,dr\,d\theta = \frac{M^2 - \pi R^2}{2\pi} \quad (14)$$

$$d_{toCH}^2 = \int_0^{\sqrt{\frac{M^2-\pi R^2}{c\pi}}} \int_0^{2\pi} r^2 \rho(r,\theta)\,r\,dr\,d\theta = \frac{M^2 - \pi R^2}{2\pi c} \quad (15)$$

Where c denoting the number of the clusters. The energy total dissipated in a network is:

$$E_{total} = sE_{Exclu} + (N-s)E_{non\_Exclu} \quad (16)$$

Where *s* is the number of the excluded nodes.

Using the Eq.11 to Eq. 16 the expected value of the energy dissipated in the network is calculated as follows:

$$E_{total} = ls\left[E_{elec} + \epsilon_{fs}\frac{R^2}{2}\right] + l(N-s)\left[NE_{elec} + (N-s)E_{DA} + c\epsilon_{mp}\left(\frac{M^2-\pi R^2}{2\pi}\right)^2\right] + l(N-s)^2\left[E_{elec} + \epsilon_{fs}\frac{M^2-\pi R^2}{2\pi c}\right] \quad (17)$$

The optimal number of clusters can be found by letting $\frac{\partial E_{total}}{\partial c} = 0$

$$C_{opt} = d_0 \sqrt{\frac{2\pi(N-s)}{M^2 - \pi R^2}} \quad (18)$$

Where $d_0 = \sqrt{\frac{\epsilon_{fs}}{\epsilon_{mp}}}$ is the distance threshold for swapping amplification models and R must be less the threshold $R_o$, where $R_o < \frac{M}{\sqrt{\pi}}$.

The different forms of the $E_{total}$ calculation will lead to different optimal $c_{opt}$ settings depending on the values of, $R$ and $s$. The optimal probability for becoming a cluster-head can also be computed as $P_{opt} = \frac{C_{opt}}{N-s}$

In Figure 3, we show the average energy consumption by each sensor node against varying numbers of clusters for different values of number of excluded nodes *s* and threshold distance *R* from base station.

While the number of cluster increases, the total energy starts to decrease and reaches a minimum for clusters number comprised between 10 and 18 depending on the value of *s* and *R*. However, it is clearly shown that when s increases, the energy consumption decreases and turns between 4.069 J and 1.473 J. These results are coincided with our conception and our goals. In the next section we have evaluate these results by computer simulation the network in Matlab.

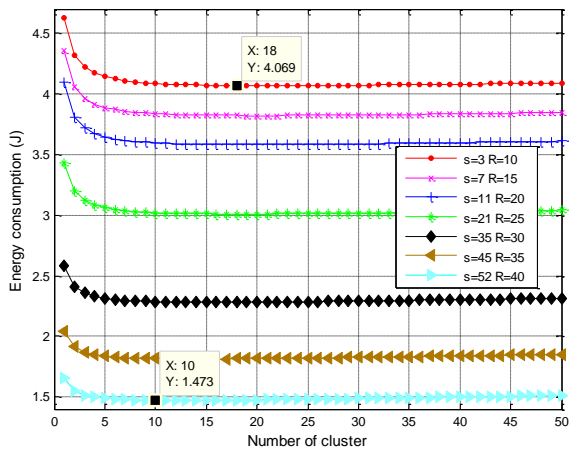

Fig. 3   Variation of energy consumption for different values of *R* and *s* depending on clusters number *c*.

## VIII. SIMULATION RESULTS

In this section, we simulate the performance of ATDEEC protocol under different scenarios using MATLAB. We consider a model illustrate in the figure 2 with N = 100 nodes randomly distributed in a 100m × 100m field. To compare the performance of ATDEEC with TDEEC protocol, we ignore the effect caused by signal collision and interference in the wireless channel. The radio parameters used in our simulations are shown in Table1.

TABLE I. ENERGY MODEL PARAMETERS

| Parameter | Value |
|---|---|
| Initial Node Energy | 0.5J |
| N | 100 |
| $E_{elec}$ | 50 nJ/bit |
| $E_{DA}$ | 5 pJ/bit |
| $\epsilon_{fs}$ | 10 pJ/bit/m$^2$ |
| $\epsilon_{mp}$ | 0.0013 J/bit/m$^4$ |
| $d_0$ | 87 m |
| L | 4000 Bytes |
| Rounds | 8000 |

We define two performance metrics to evaluate our protocol as: First Node Dies (FND), or stability period and Last Node Dies (LND), or instability period.

First, we present an empirical result for the optimal number of cluster-head $C_{opt}$ and optimal threshold distance to the base station for our ATDEEC protocol shown in Figure 4. The number of cluster-heads decreases from 10 to 45 meters. This figure reveals that although the cluster-heads decreases from 5 to 17, the FND improves significantly and has a maximum value at 20 meters. Beyond this value, the curve starts descending. The optimality of $C_{opt}$ lies around 17 cluster-heads for our setup. This result can be interpreted by when the threshold distance R start to increase, the closer nodes to the base station consume less energy, because they send data directly to it. However, when this distance increases the nodes become farther away and consume more energy.

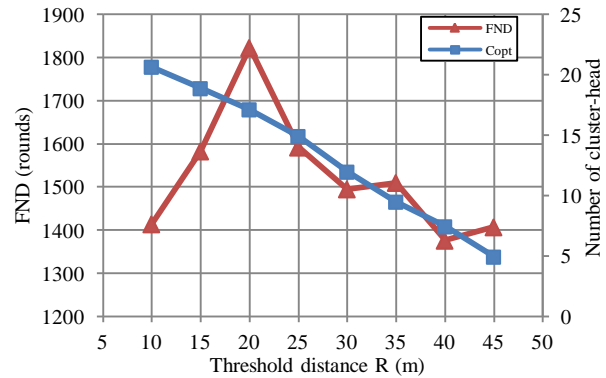

Fig. 4   FND and $C_{opt}$ vs Threshold distance R

On other hand, we study three other performance metrics such as, the number of live nodes per round, energy residual and number of message packets for both ATDEEC and TDEEC protocols. The simulation results are discussed below.

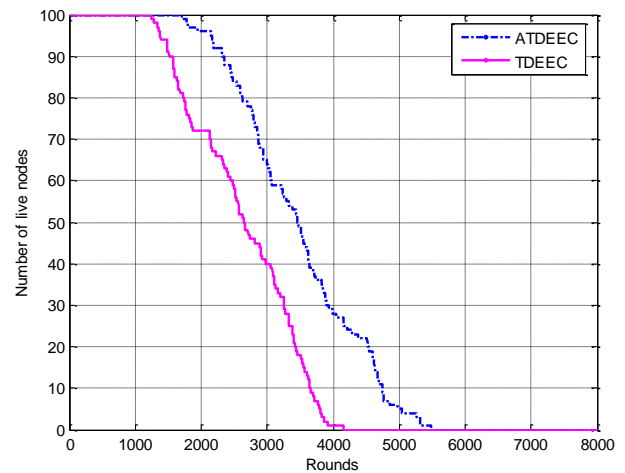

Fig. 5   Life time ATDEEC and TDEEC comparison

Figure 5 shows the network lifetime of ATDEEC and TDEEC for threshold distance equal to 20m. Since the TDEEC protocol is designed to be robust with respect to a heterogeneous network, we test the performance of ATDEEC against these criteria. Based on our experimental results, we conclude that ATDEEC has a superior stability period life time performance compared with TDEEC by an increase with 25% as shown in this same figure.

In the Figure 6, we emphasis our discussion on how each node consumes its own residual energy in the network. This energy is calculated during the network operation, by observing the variation of energy levels between the nodes at each round. The total initial energy of the network is 90 J which decreases linearly up to 3000 rounds and after that there is a difference from the round where first node dies in respect to them. Energy residual per round for ATDEEC is more as compared to TDEEC.

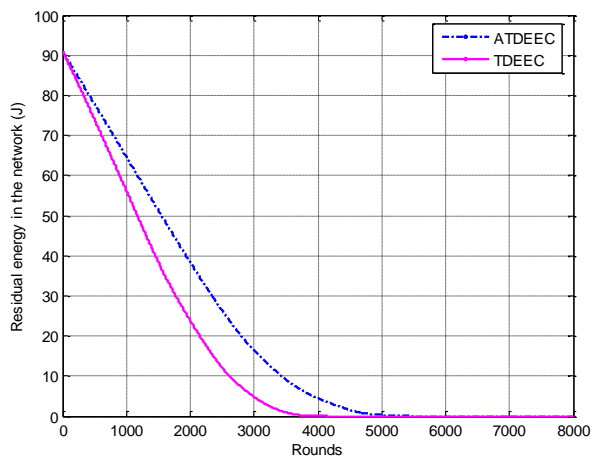

Fig. 6   Total residual energy over rounds TDEEC and TDEEC

Referred to figure 7, it show clearly that proposed approach provide a better throughput compared to TDEEC protocol, this increase is justified by the life time enhancement which give the improved ATDEEC protocol.

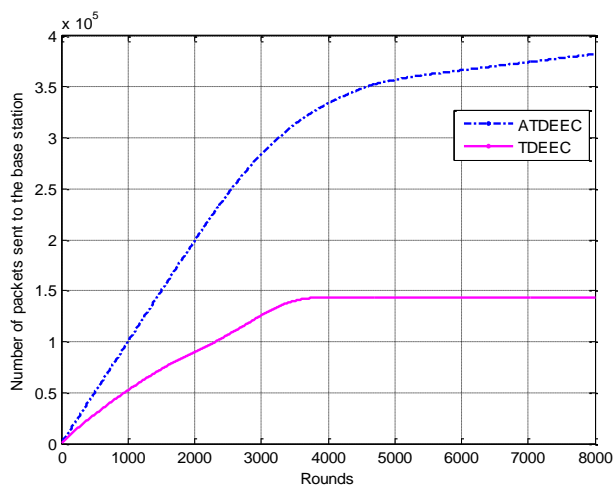

Fig. 7   Performance of the protocols

Generally, we can illustrate the increase of the proposed protocol in the figure 8. It's noted that the throughput increases twice as much than TDEEC due to its energy efficiency. Whereas, ATDEEC outperforms the FND of TDEEC by 25% and by 46% for LND.

## IX. CONCLUSION AND FUTURE WORK

In this paper, an energy efficient protocol ATDEEC has been proposed to solve the problem of the closest nodes to the base station which were consumed more energy in data traffics.

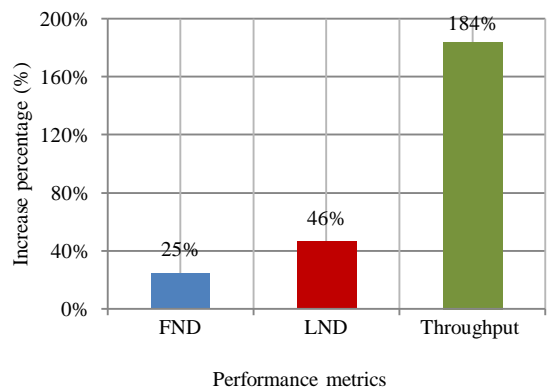

Fig. 8   Performance metrics of the ATDEEC protocol

The simulation result by Matlab, demonstrate the ability of developed algorithm to prolong the network lifetime significantly and increase the number of packet messages received by the base station. In the future work we'll evaluate this approach by the real-time performances and simulate it by adequate simulator software.